\documentclass[english,conference]{IEEEtran}
\usepackage[T1]{fontenc}
\usepackage[latin9]{inputenc}
\usepackage{array}
\usepackage{multirow}
\usepackage{amstext}
\usepackage{graphicx}
\usepackage{xcolor}
\usepackage{url}
\usepackage{tabularx}
\makeatletter
\usepackage{blindtext}
\usepackage{enumitem}

\usepackage[T1]{fontenc}
\usepackage[latin9]{inputenc}
\usepackage{array}

\usepackage{xcolor}

\usepackage{graphicx}
\usepackage[compatibility=off]{caption}
\usepackage{subcaption}
\providecommand{\tabularnewline}{\\}

\setkeys{Gin}{width=\linewidth}

\makeatother

\usepackage{babel}

\usepackage{geometry}
\begin{document}
	\begin{flushleft}
		{\Large
			\textbf\newline{Discovery, Retrieval, and Analysis  of the `Star Wars' Botnet in Twitter} 
		}
		\newline
		\\
		Juan Echeverria\textsuperscript{*}, 
		Shi Zhou
		
		\bigskip
		Department of Computer Science\\University College London (UCL), United Kingdom
		\\
		* J.EcheverriaGuzman@ucl.ac.uk
		
\end{flushleft}

\begin{abstract}

	It is known that many Twitter users are bots, which are  accounts controlled and sometimes created by computers. Twitter bots can send spam tweets, manipulate public opinion and be used for online fraud.  
	Here we report the discovery, retrieval, and analysis of the `Star Wars' botnet in Twitter, which consists of more than 350,000 bots tweeting random quotations exclusively from Star Wars novels. 
		The botnet contains a single type of bot, showing exactly the same properties throughout the botnet. It is unusually large,  many times larger than other available datasets.  
	It  provides a valuable source of ground truth  for research on Twitter bots.  
	We analysed and revealed rich details on how the botnet was designed and created. As of this writing, the Star Wars bots are still alive in Twitter. 
	They have survived since their creation in 2013, despite the increasing efforts in recent years to detect and remove Twitter bots.
	We also reflect on the `unconventional' way in which we discovered the Star Wars bots, and discuss the current problems and future challenges  of Twitter bot detection.

\end{abstract}

\section{Introduction}

Twitter plays an increasingly important role in modern society.   
There are more than 313 million active users, and about 500 million tweets are created everyday \cite{about_twitter}.    
Twitter has been the subject of intensive study in recent years, including the graph structure of Twitter users \cite{kwak_what_2010}   and  ways to measure the influence of a Twitter user \cite{cha_measuring_2010}.  A popular research trend is to analyse the real-time stream of tweets as data source to detect and predict  events,  such as outbreak of epidemics \cite{lampos_flu_2010},  election results \cite{tumasjan_predicting_2010-1}, earthquakes and typhoons \cite{sakaki_earthquake_2010}. 
 
\subsection{Twitter bots and botnet}

There are a lot of fake Twitter user accounts, called  bots \cite{Bot_tutorial}, that are used and controlled by computers and algorithms, instead of real people. 
A botnet is a collection of bots  that have the same properties  and are controlled  by the same 'botmaster'    \cite{boshmaf_socialbot_2011-1}.

\subsection{Threats of Twitter bots } \label{section:threats}

Some Twitter bots  have benign purposes, for example  automated tweet alert of a new research article or weather forecast. 
%
The subject of our research is the large number of bots that have non-benign purposes, such as the followings.

\paragraph{Spamming}
Spammer bots can send a large amount of unsolicited content to other users. The most common objective of spam is getting users to click on advertising links with questionable value \cite{grier_spam:_2010}, or propagate computer viruses and other malware.

\paragraph{Fake trending topics} If  bots are able to pass as humans through Twitter's filters, they would be counted by Twitter for choosing trending topics and hashtags. This would allow the bots to create fake trending topics that are not actually being popular in Twitter. 

\paragraph{Opinion manipulation}
A large group of bots   can misrepresent public opinion \cite{forelle_political_2015}. If the bots are not detected in time, they could tweet like real users, but coordinated centrally around a specific topic. They could all post positive or negative tweets skewing metrics used by companies and researchers to track opinions on that topics.

\paragraph{Astroturfing}
Bots can orchestrate a campaign to create a fake sense of agreement among Twitter users \cite{abokhodair_dissecting_2015,ratkiewicz_truthy:_2011}, where they  mask the sponsor of the message, making it seem like it originates from the community itself.

\paragraph{Fake Followers}
Fake followers can be bought or sold online \cite{yang_analyzing_2012}. After receiving payment from a user, the botmaster of a botnet  can instruct its bots to follow that user.   Fake followers could make a user seem more important than it is  \cite{cha_measuring_2010,messias_you_2013,freitas_reverse_2014}. 
One would expect that fake followers should try to appear like real users   \cite{gratton_follow_2012,grier_spam:_2010}, however people rarely verify whether someone's followers are human or bots.

\paragraph{Streaming API contamination}

Many research works rely on analysing tweet data returned by Twitter's streaming API.  It is reported \cite{morstatter_can_2016} that the  API is susceptible to an attack by bots, where bots can time their tweets in such a way that their tweets can be included in the API with a probability higher than the expected  1\%, up to as high as 82\%.

\subsection{Twitter bots detection}

It has been reported that Twitter has been  identifying and removing suspicious users, many of which are spammer bots \cite{thomas_suspended_2011}. 
Twitter does not disclose their bot data.
One may wonder whether Twitter  removes as many bots as it possibly can. 

Researchers have  proposed a series of methods to detect  Twitter bots \cite{subrahmanian_darpa_2016,thomas_trafficking_2013,thomas_suspended_2011}.  
For example,  \cite{wang_detecting_2010}  used the Levensthein distance between tweets of each user to identify bots;  
\cite{zafarani_10_2015} and \cite{lee_early_2014}  aimed to classify bots  quickly with minimum information;
\cite{abokhodair_dissecting_2015}  discovered a botnet  of 130 bots. 

It is recognised that a major challenge to the research on Twitter bots  is a lack of ground truth data \cite{subrahmanian_darpa_2016}.  So far,  datasets that have been made publicly available by researchers are small, with the largest dataset contains only a few thousands of bots. All these datasets  contain many types of bots from 
different botnets. Many identified bots have  been suspended  and therefore  disappeared from Twitter.

\subsection{Contributions of this paper}

Here, we report the discovery of the  'Star Wars' botnet in Twitter. This dataset is unique and valuable for a number of reasons. 
First, it is unusually large, containing more than 350,000 bots; second, it is the first dataset that only contains one type of bots  from a single  botnet and therefore showing the same properties; third, the identified bots, at the time of this publication, are still alive in Twitter, allowing researchers to collect, monitor and study them; and finally, the Star Wars bots exhibit some interesting  properties that have not been reported before. New knowledge learned from this botnet has already allowed us to discover another botnet with more than half a million bots, which will be reported in a separate paper.

\section{Discovery of the Star Wars  bots}

The Star Wars bots were discovered by chance when we studied the location of tweets created by a 1\% random sample of Twitter users. 

%
%
 
\subsection{Data of 1\% random Twitter users}

There have been efforts to collect complete datasets from Twitter \cite{abisheva_who_2014,ellis_equality_2016}. Most studies, however, have relied on sampled datasets \cite{krishnamurthy_few_2008,stutzbach_unbiased_2009}. 
We can use Twitter stream API to get samples of real-time tweets \cite{rajadesingan_sarcasm_2015}.  Or we can obtain a sample of Twitter users and then use Twitter API to retrieve the latest 3,200 tweets of each of the users. Twitter users can be sampled in various ways, such as random walk and breadth-first search   \cite{leskovec_sampling_2006,krishnamurthy_few_2008} on the Twitter user graph.

Twitter users can also be collected by random uniform sampling  \cite{wang_unbiased_2010,gabielkov_complete_2012}.
While each user  chooses its own  username, Twitter assigns each user a unique 32-bit ID number\footnote{Twitter has extended the user ID space to  64 bits, which has no effect on this work.}. 
In early 2015, we randomly chose user IDs with a uniform 1\% probability in the ID space ranging from 1 to $2^{32}$. We  retrieved  the user profile for each valid user ID. In this work, we  considered only English-speaking users, so we filtered out any user whose declared interface language was not English.

We collected 6 million random (English speaking) Twitter users, which are called the \emph{1\% random users} dataset.

\subsection{Abnormal distribution of tweet locations }\label{section:abnormalmap}

Different from the registered location in its  profile,  a Twitter user can choose to tag each tweet with the latitude and longitude coordinates where the tweet is created. Change of tweet locations reflects a user's mobility. 

\begin{figure*}[t]
	\centering
		\includegraphics [width=15cm] {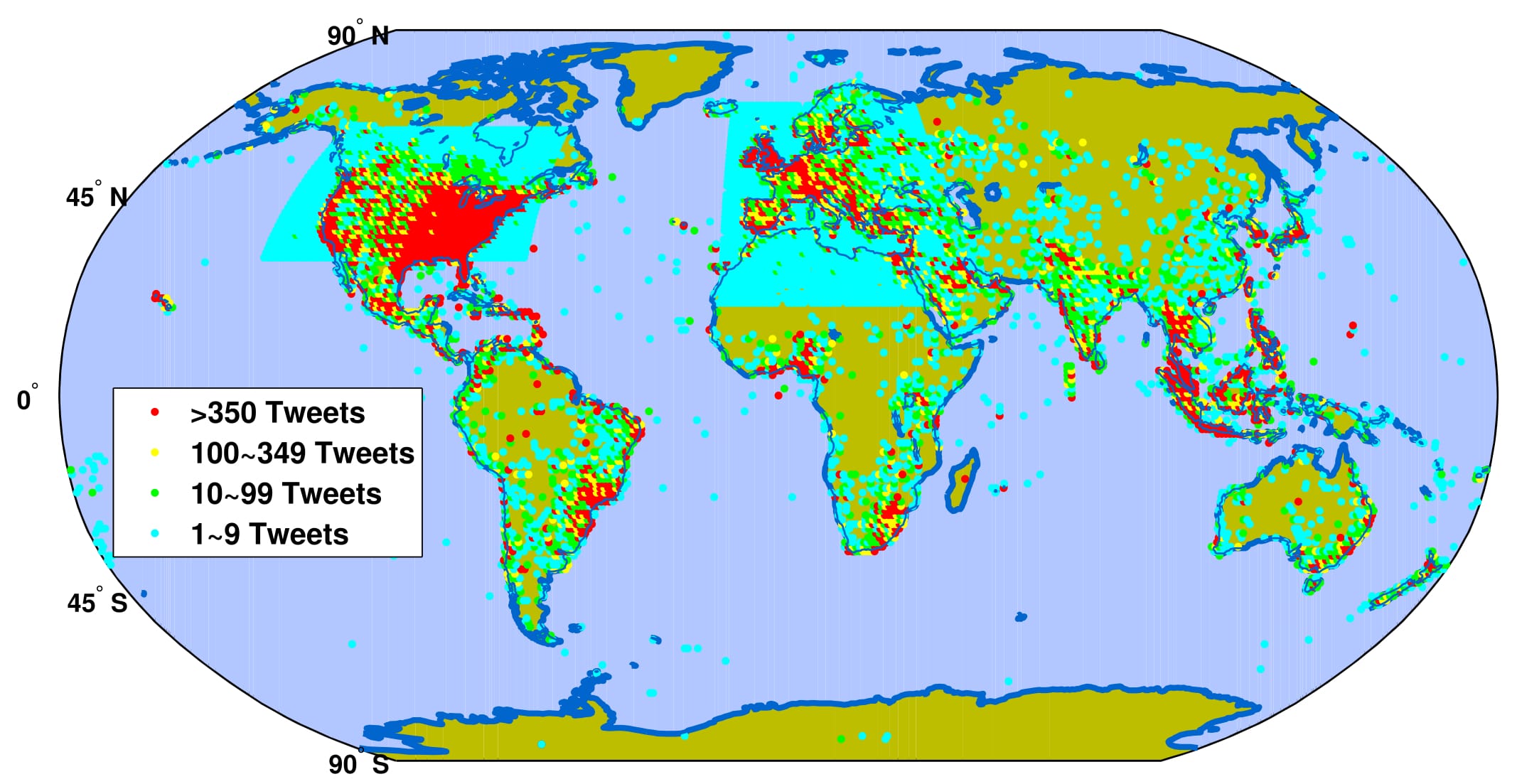}
		\caption*{(a) World}
	
	\begin{subfigure}[b]{8cm}
		\includegraphics [width=8cm] {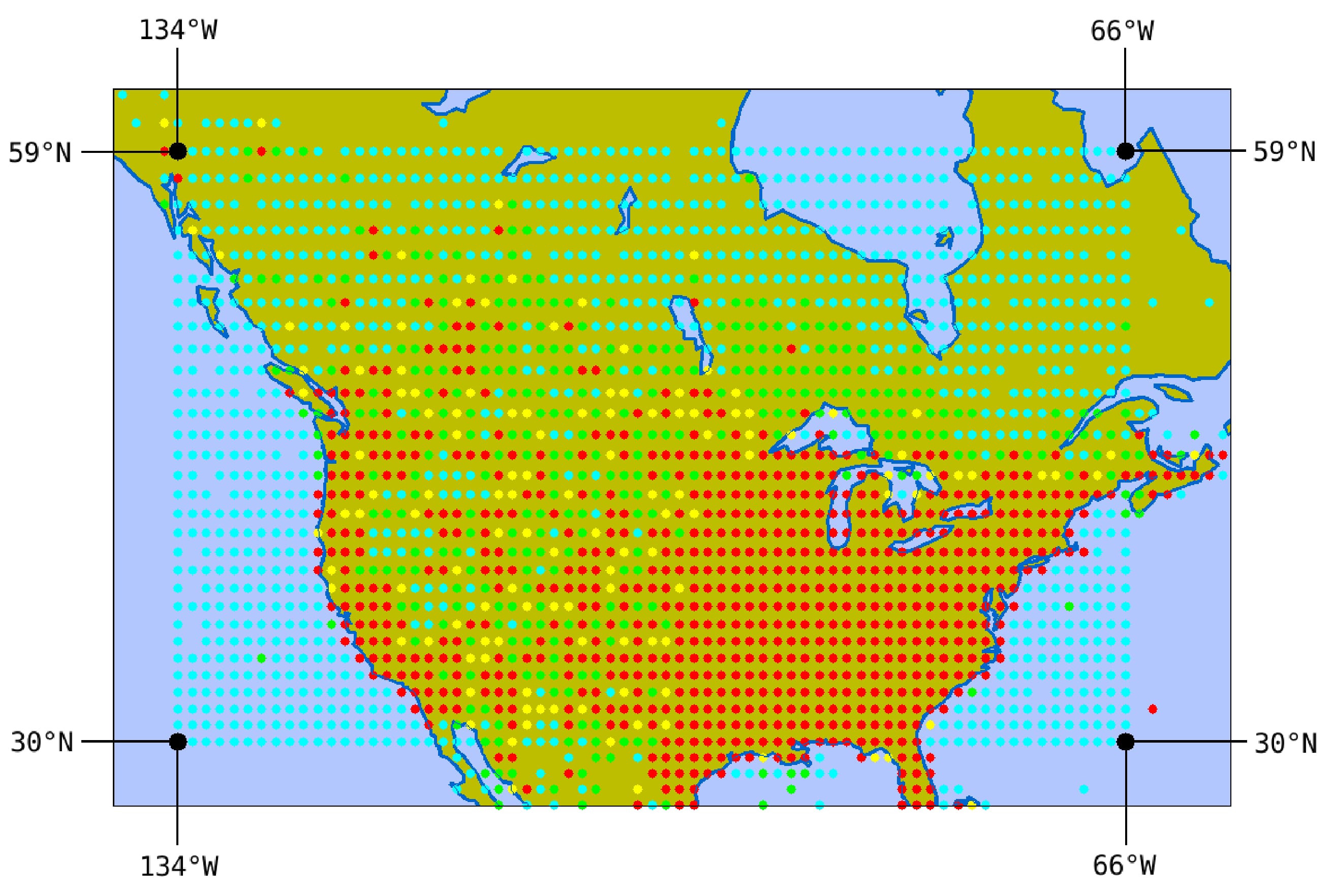}
		\caption*{(b) North America}
	\end{subfigure} 
	\hspace{5mm}
	\begin{subfigure}[b]{6cm}
		\includegraphics  [width=6cm]{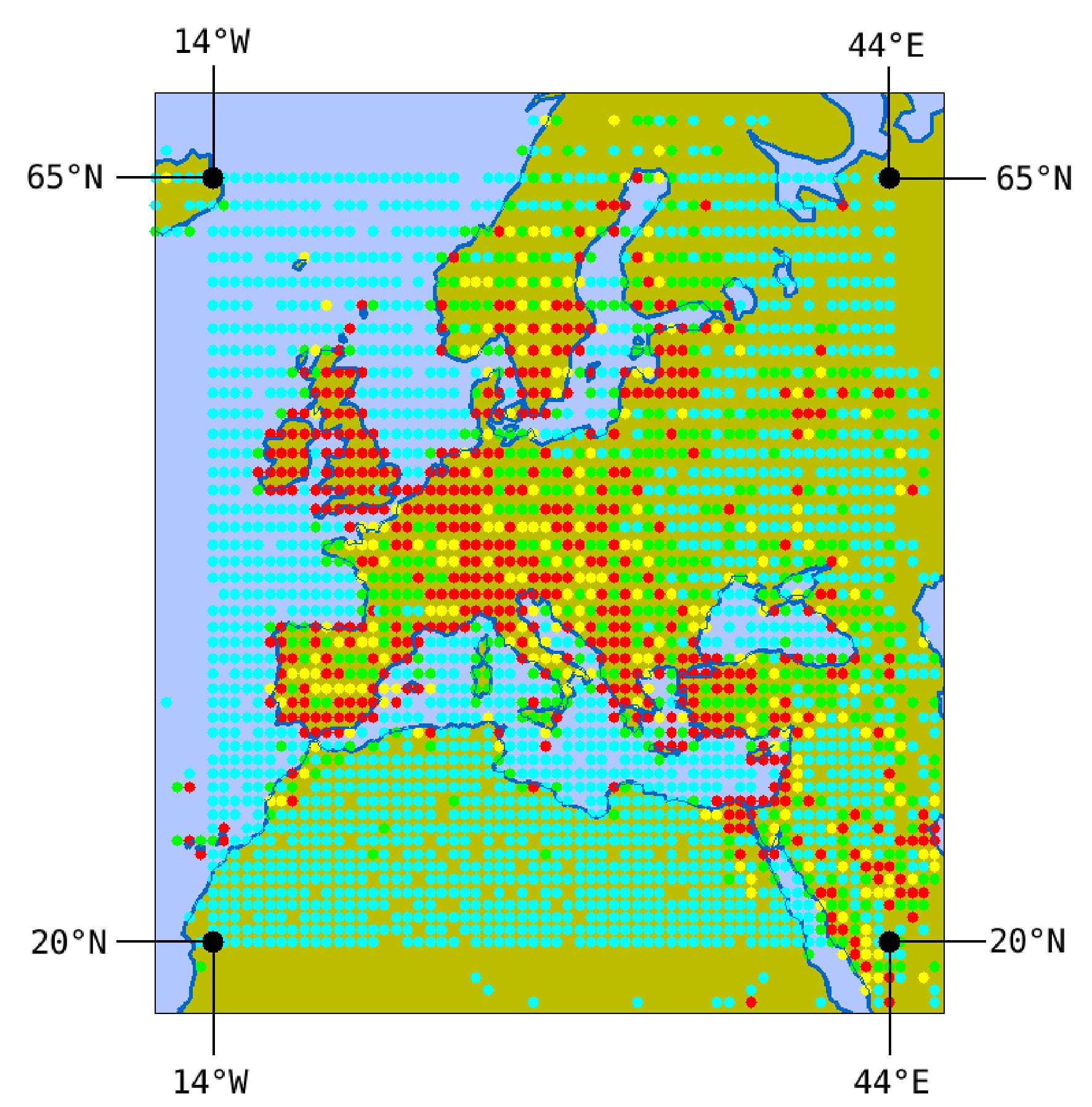}
		\caption*{(c) Europe}
	\end{subfigure} 
	\caption{Distribution of  geographic locations of tweets created by 1\% random Twitter users. The map is divided into 
		cells of 1 latitude and 1 longitude. The number of tweets in each cell  is coded with colours.  } 
	 \vspace{-20pt}	\label{fig: RndUsersMapAll}
\end{figure*}

The 1\% random users had 843 million tweets, of which 20 million   had a location tag.  
Figure \ref{fig: RndUsersMapAll} shows the number of tweets located in each geographic cell of $1^o$\,longitude width and $1^o$\,latitude height.  
While the distribution of tweet locations is in general coincident with the distribution of Twitter users population, there is an anomaly in the form of two rectangular areas over North America and Europe that  are solidly filled with blue colour cells (with 1 - 9 tweets) over large uninhabited areas such as seas, deserts and frozen lands. 
Outside the two rectangles, there are only a few blue colour cells  on such uninhabited areas.

The two rectangles have sharp corners and straight borders that are parallel to the latitude and longitude lines. 
These are distinct characteristics of computer automation. 
%
%

Our instinct was that these uniform, low-density blankets of tweets  over uninhabited area were likely created by Twitter bots. Computer programs can easily tag  tweets with randomly generated latitude and longitude values in the ranges covering the borders of North America and Europe/North Africa. 
%
%
%
A possible explanation is that the bots might try to pretend that they tweet from North America and Europe where Twitter is  popular.

\subsection{Tweets of random quotations from  Star Wars novels }

 %
%
%
There are 23,820  tweets located in the blue-colour cells inside the two rectangles. We manually checked the text of these tweets. We found out that many of these tweets were randomly quoted from Star Wars novels. It is known that Twitter bots often quote from books or online sources \cite{gratton_follow_2012}. We identified that at least 11 different Star Wars novels are used as the source of the tweets. 
 Here is an example tweet: 
\begin{quotation}
	\textit{Luke's answer was to put on an extra burst of speed. There were only ten meters \#separating them now. If he could cover t}
\end{quotation}
\begin{figure}
	\centering
	\includegraphics{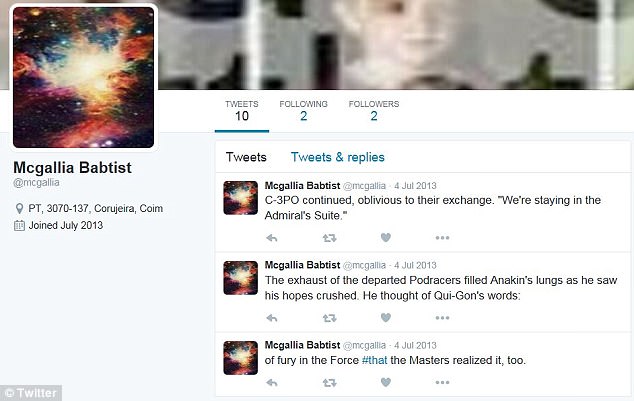}
	\caption{A Star Wars bot as it appears on Twitter}
	\label{fig: botexample}
\end{figure}
This tweet is quoted  from the book \emph {Star Wars: Choices of One}, where Luke Skywalker is a distinct character. 
Some tweets started or ended with a truncated word.  Some tweets have a hashtag or a hash symbol inserted at a random place. Figure \ref{fig: botexample} shows how a Star Wars  bot looks in the Twitter homepage.

\begin{figure} 
	\centering
	\includegraphics[width=\columnwidth]{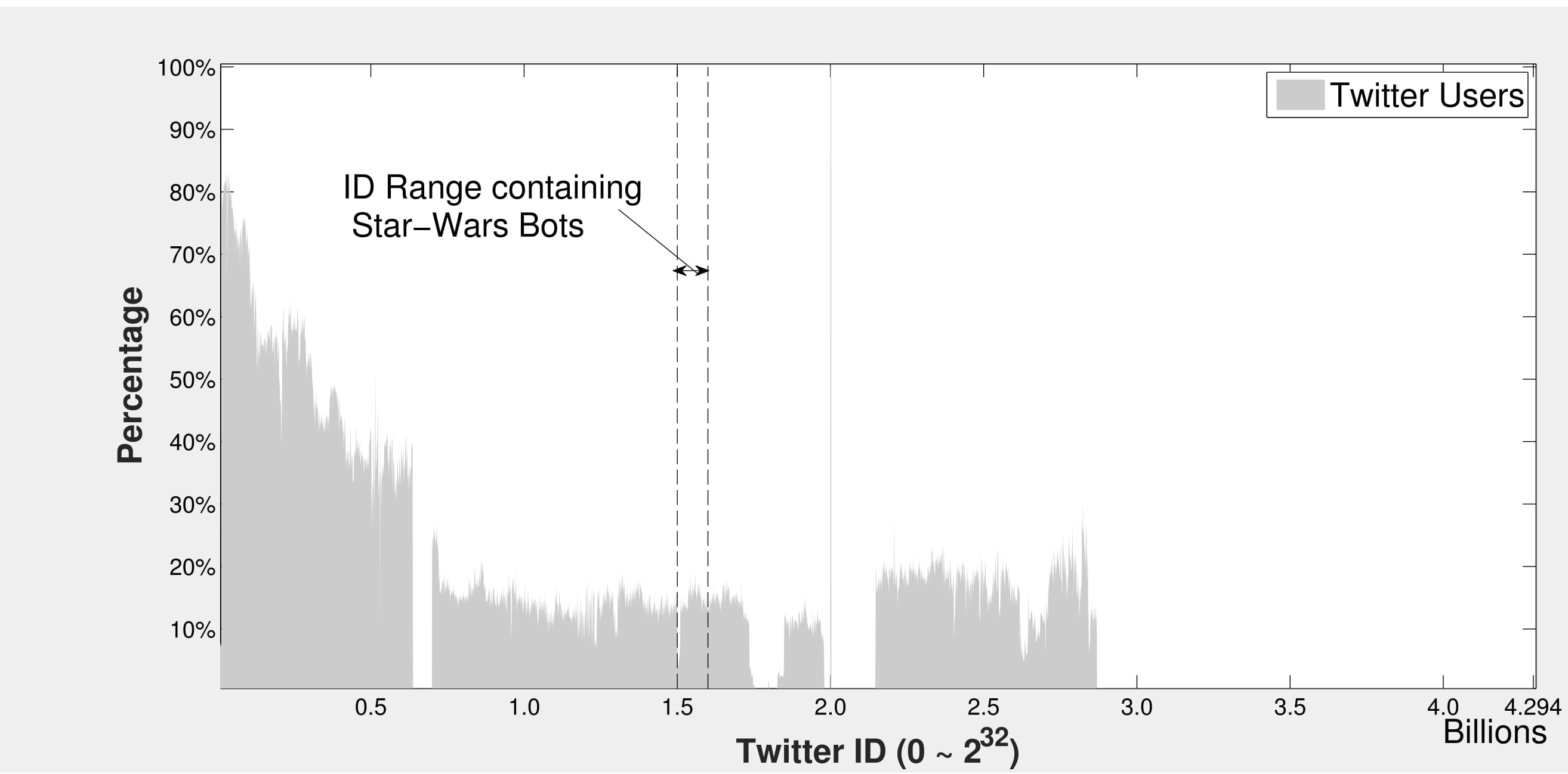}
	\caption*{(a)} 
	
	\vspace{5mm}
	
	\includegraphics[width=\columnwidth]{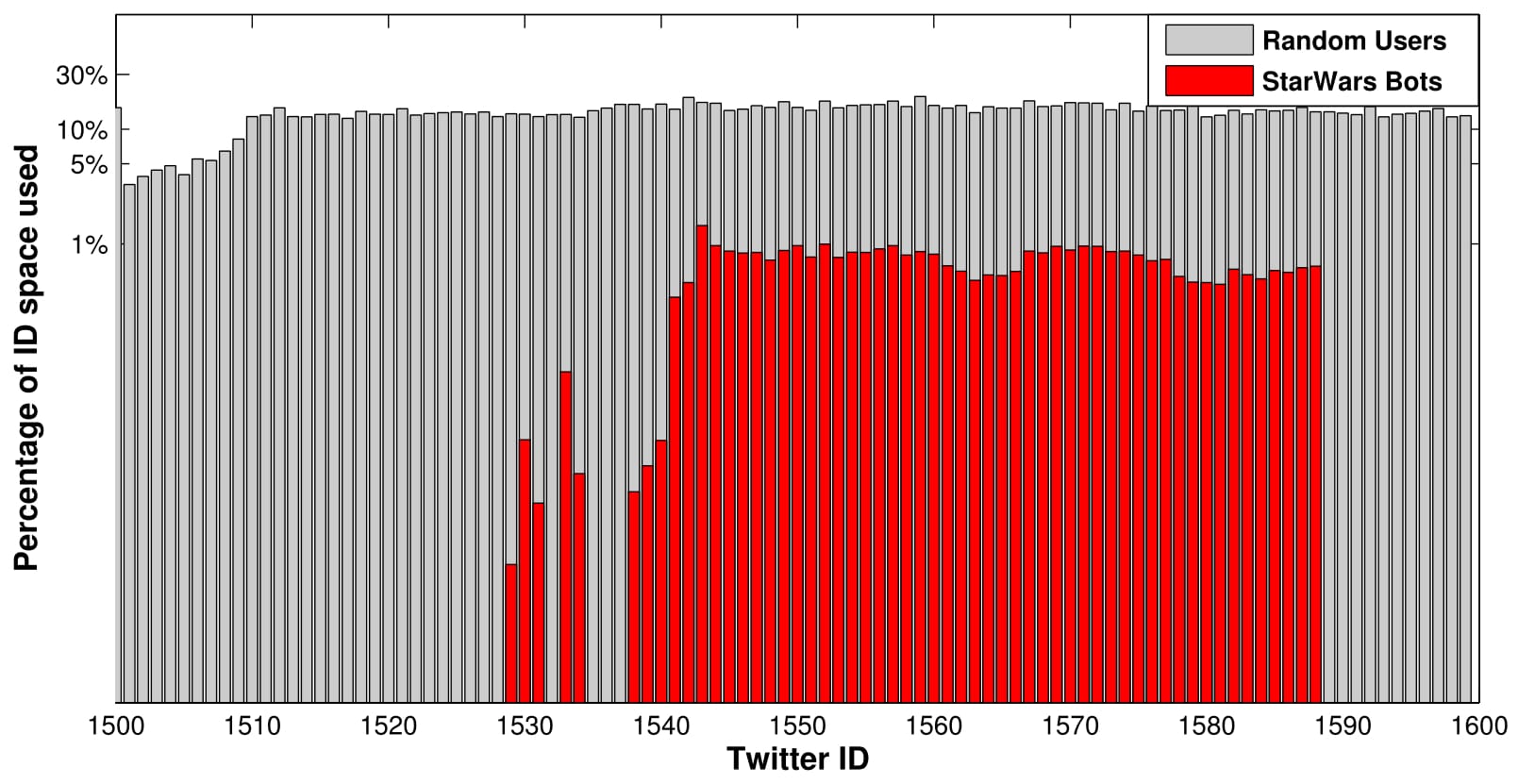}
	\caption*{(b)}

	\caption{Distribution of Twitter user IDs. (a) Percentage of  user IDs that are in use in each bin of 1 million IDs.  (b) Distribution of user IDs of the Star Wars bots (in red colour), where the y axis is shown in log scale.	}\label{fig : ID_distribution}
	
\end{figure}

\subsection{Definition of Star Wars bots}\label{section:SWbots}

The  tweets in the blue-colour cells are created by 3,244 Twitter users. 
%
%
{Upon} further examination, we found that these suspicious users also show \emph{all} of the following  properties. 
\begin{itemize}
	\item  They {\em only} tweet random quotations from the Star Wars novels. Each tweet contains  one quotation, often with incomplete sentences or broken words at the beginning or at the end. 
	\item The only extra text that is sometimes inserted in such a tweet is either (1) one of the special hashtags associated with attracting followers, such as \textit{\#teamfollowback} and \textit{\#followme}; or (2) the hash symbol \textit{\#}   inserted in front of a randomly chosen word (including the  stop words, like  "the" and "in")   to form a hashtag.
	\item They have created no more than 11 tweets in their lifetime.	
	They never retweet; and they never mention any other Twitter user. 	
	
	\item They have no more than 10 followers, and no more than 31 friends. 

	\item They only choose `Twitter for Windows Phone' as the source of their tweets. 
		\item {Their Twitter user IDs are densely concentrated within a narrow space between $ 1.5\times10^9$ and $1.6\times10^9$.  See Figure \ref{fig : ID_distribution}.}
\end{itemize}

%

These distinct, unusual properties convinced us that we have discovered a new type of Twitter bots, the Star Wars bots, which are defined as Twitter users that exhibit {\em all} properties in the above list.

\section{Retrieval of the Star Wars botnet }

In the above section we discovered the Star Wars bots manually from the 1\% random users. In this section we will use an automatic classifier to retrieve all  Star Wars bots that form the whole Star Wars botnet  based solely on textual features of tweets created by a user.

 \begin{table}
	\center
	\begin{tabular}{|c|c|c|c|}
		\hline 
		Classifier results& Star Wars Bots & Real Users & Total\tabularnewline
		\hline 
		\hline 
		Correct & 2996 & 8987 & 11983\tabularnewline
		\hline 
		Incorrect & 4 & 13 & 17\tabularnewline
		\hline 
		Accuracy & 99.8\% & 99.8\% & 99.8\%\tabularnewline
		\hline 
	\end{tabular}
	
	\protect\caption{\label{tab:Naive-Bayes-Classifierconfusionmatrix}The  Confusion Matrix 
	}
\end{table}

 \begin{figure}
 	\centering
 	
 	\includegraphics[width=\columnwidth]{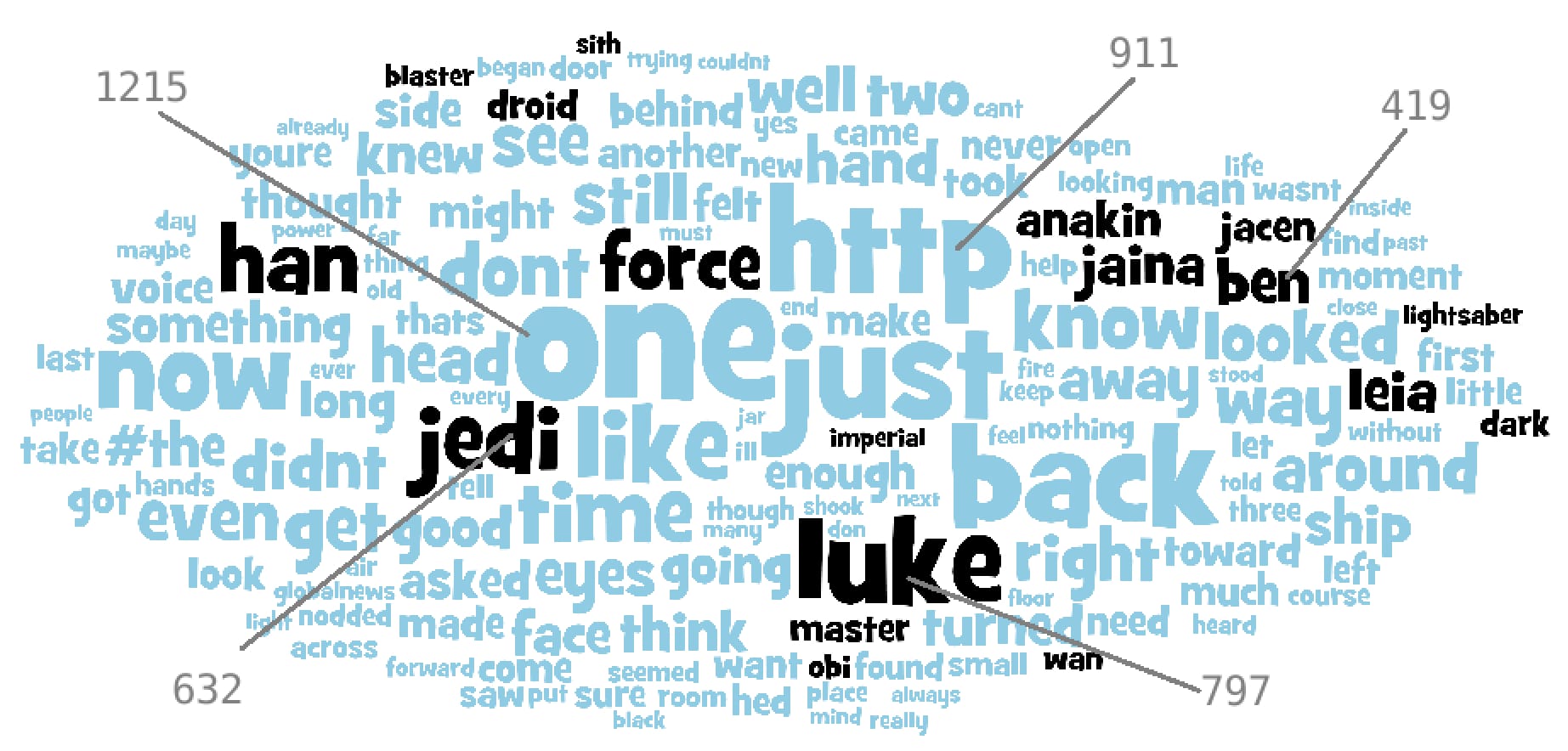}
 	\caption*{(a)}

 	\includegraphics[width=\columnwidth]{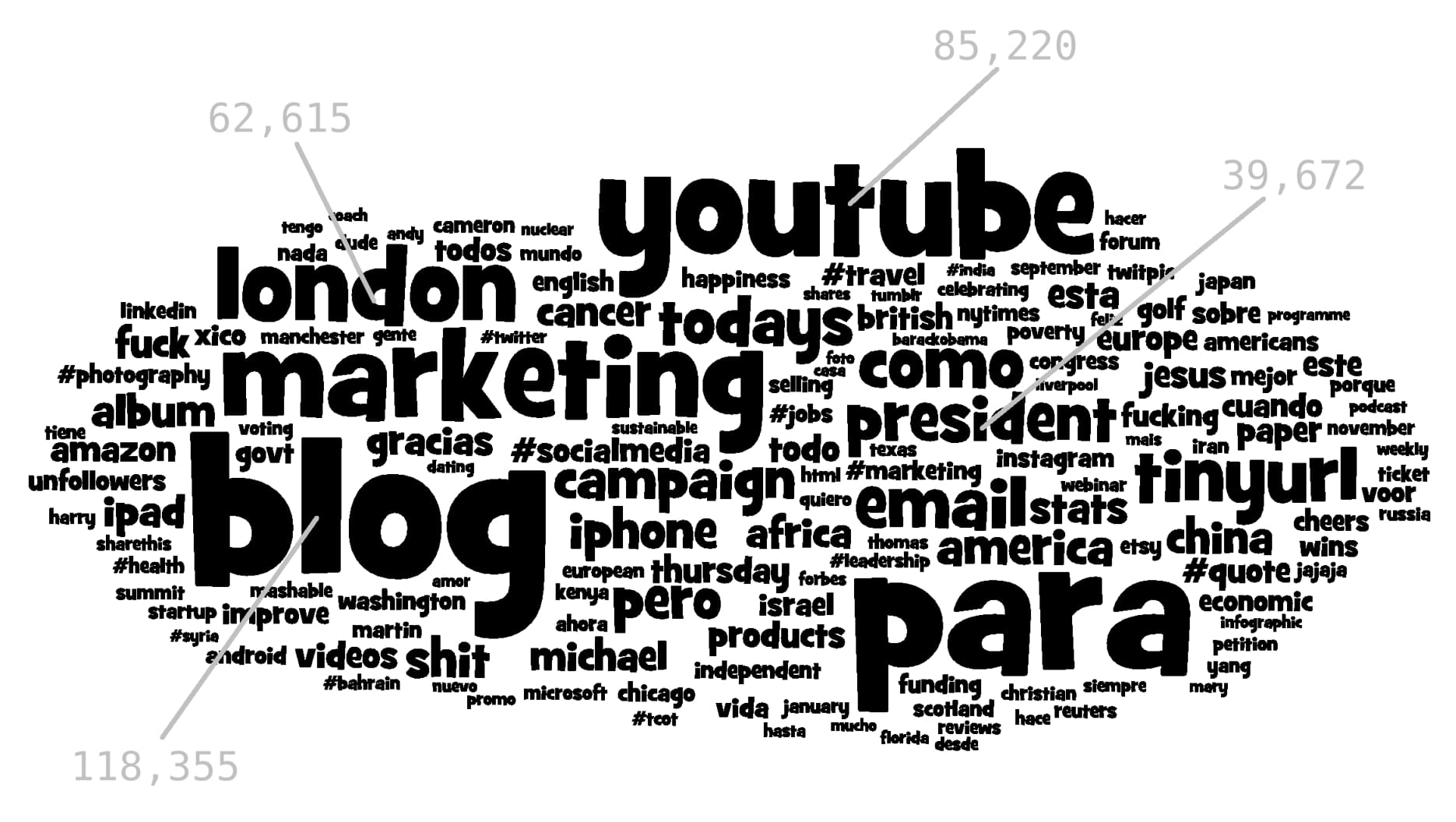}
 	\caption*{(b)}
 	
 	\caption{Word cloud representations of the  most frequently tweeted words by (a)  Star Wars bots and (b)  real   users. The font size of a word indicates the frequency of the word's appearance in tweets. The frequency number of some selected words are shown.  
 	}\label{fig : wordcloud}
 \end{figure}

To contrast against the Star Wars bots, we collect a dataset of 9,000
random real users. 
Starting from a known real user, we collected the friends that it follows, and then collected the friends of its friends. We collected four levels of friends using breadth-first search. We got two million users. We  randomly selected 9k English-speaking users. These users  are mostly real users, because a real user's friends are usually also real users.

We collected all tweets of 3,000 of the Star Wars bots  that are manually tagged in the above section, and all tweets of the 9,000  real users. We  remove  stop words and  non alphabetical characters from the tweets except the \#  symbol. 

We  obtained 30,000 words that were most frequently tweeted   by the bots, and 50,000 by the real users.
Based on these words, we created a word count vector for each bot and real user. 

We  used this training dataset to train a Naive Bayes classifier. 
We tested the classifier with 10-fold cross validation. The classifier achieved  $>$99\% for  precision, recall and F-measure on each of the folds. The confusion matrix  is shown in table \ref{tab:Naive-Bayes-Classifierconfusionmatrix}.

As a sanity test, we also created a balanced training dataset of 9k real users and 9k bots (by adding two copies of the bots). Training with this dataset produced near identical results. 

It is evident that this  basic machine learning classifier is sufficient for this task. This is because tweets created by the Star Wars bots show textual features that are distinct from those created by real users, as illustrated in Figure \ref{fig : wordcloud}.

 	\begin{table*}
		\center
		\begin{tabular}{ rrr}
			\hline 
			\centering{\bf Properties} & \centering{\bf Random  users} & \centering{\bf Star\,Wars bots} \tabularnewline
			\hline 
			{Number of users} &\ 6,063,970 & 356,957 \tabularnewline
			\hline 
			Users with location-tagged tweets & 208,612  & 349,045 \tabularnewline
			\hline 
			\% users with  location-tagged tweets  & 3.4\% & 97.7\%\tabularnewline
			\hline 		
			{Number of tweets} &  842,670,281   & 2,422,013 \tabularnewline
			\hline 
			Location-tagged tweets & 19,777,003   & 1,209,597 \tabularnewline
			\hline 
			\% of  location-tagged tweets	 & 2.3\% &  49.9\%\tabularnewline
			\hline \hline
			{\bf Of the location-tagged tweets}\tabularnewline
			\hline
			Tweets in EU rectangle  & 3,486,239& 604,647 \tabularnewline
			\hline 
			\% of  tweets in EU rectangle	 & 17.6\% & 50.0\%\tabularnewline		
			\hline 
			Tweets in N.\,Am.\,rectangle  & 8,950,941 & 604,912 \tabularnewline
			\hline 
			\% of  tweets in N.\,Am.\,rectangle 	 & 45.2\% & 50.0\%\tabularnewline
			\hline 	
			Tweets elsewhere &  7,339,823 & 38 \tabularnewline
			\hline 
			\% of  tweets elsewhere	 &  37.1\% & 0.0\%\tabularnewline
			\hline 
			\hline		
			\centering{\bf Tweet source}	  \tabularnewline\hline
			Twitter for iPhone& 31.1\%& -\tabularnewline\hline
			
			Twitter Web client& 17.2\%& -\tabularnewline\hline
			Twitter for Android& 14.9\%& -\tabularnewline\hline
			
			Twitter for Blackberry& 6.8\%& -\tabularnewline\hline
			Mobile Web&0.99\%&-\tabularnewline\hline
			Twitter for Windows Phone&0.02\%&{100\%}\tabularnewline\hline
			
		\end{tabular}
		\protect\caption{Properties of the Star Wars bots and the 1\% random users.  
			\label{tab:property} }
		\vspace{-5mm}
	\end{table*}  
	
According to the definition of Star Wars bots, we collected the 14 million English-speaking Twitter users whose IDs were between $ 1.5 \times 10^9 $ and $1.6\times 10^9$. 
We removed users  with $>$ 11 tweets, $>$ 10 followers or $>$ 31 friends. We also removed users whose tweet source is not 'Twitter for Windows Phone'. 
From the remaining users, the classifier retrieved 356,957 Star Wars bots. 

\section{Analysis of the Star Wars botnet}

Table \ref{tab:property} shows properties of the 356,957 Star Wars bots in comparison to the 1\% random users. 
\subsection{Tweet location properties}

The Star Wars bots are retrieved by the classifier based solely on  tweet text features without tweet location information.
As shown in Table \ref{tab:property},   very few random users have location-tagged tweets.  In contrast,  97.7\% of the Star Wars bots have at least one location-tagged tweet; and almost exactly half of all tweets created by the bots have location tags.  

The 1.2 million location-tagged tweets of the Star Wars bots are distributed inside the two rectangles with an exact 50/50 split, except only 38 tweets falling outside the rectangles. 

\begin{figure} 	
	\centering
	\includegraphics[width=\columnwidth]{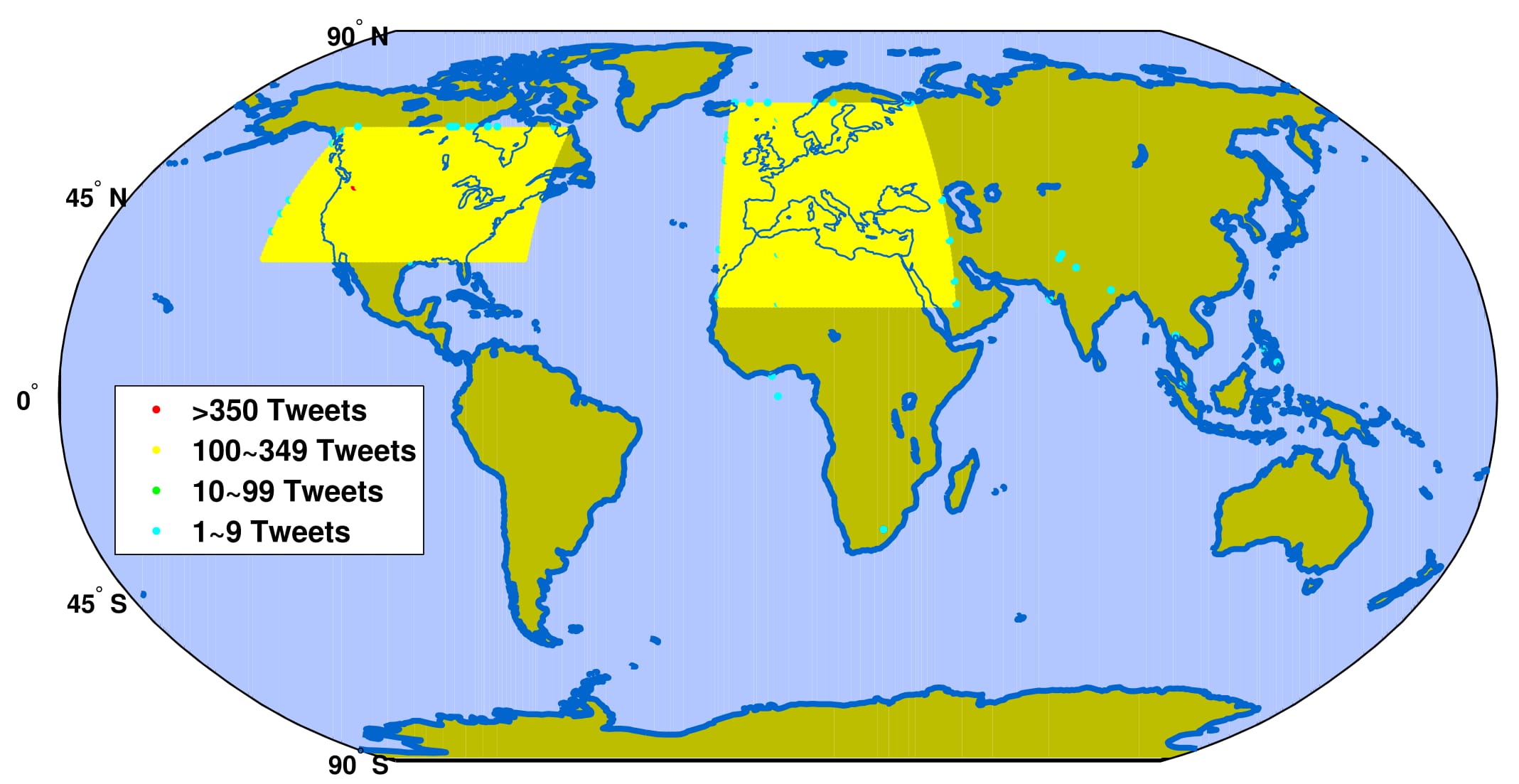}
	\caption{Geographic distribution of 1.2 million location-tagged tweets created by the Star Wars bots.}
	\label{fig: Map bots}
\end{figure}

\begin{figure*}
	\centering
	\includegraphics[width=12cm]{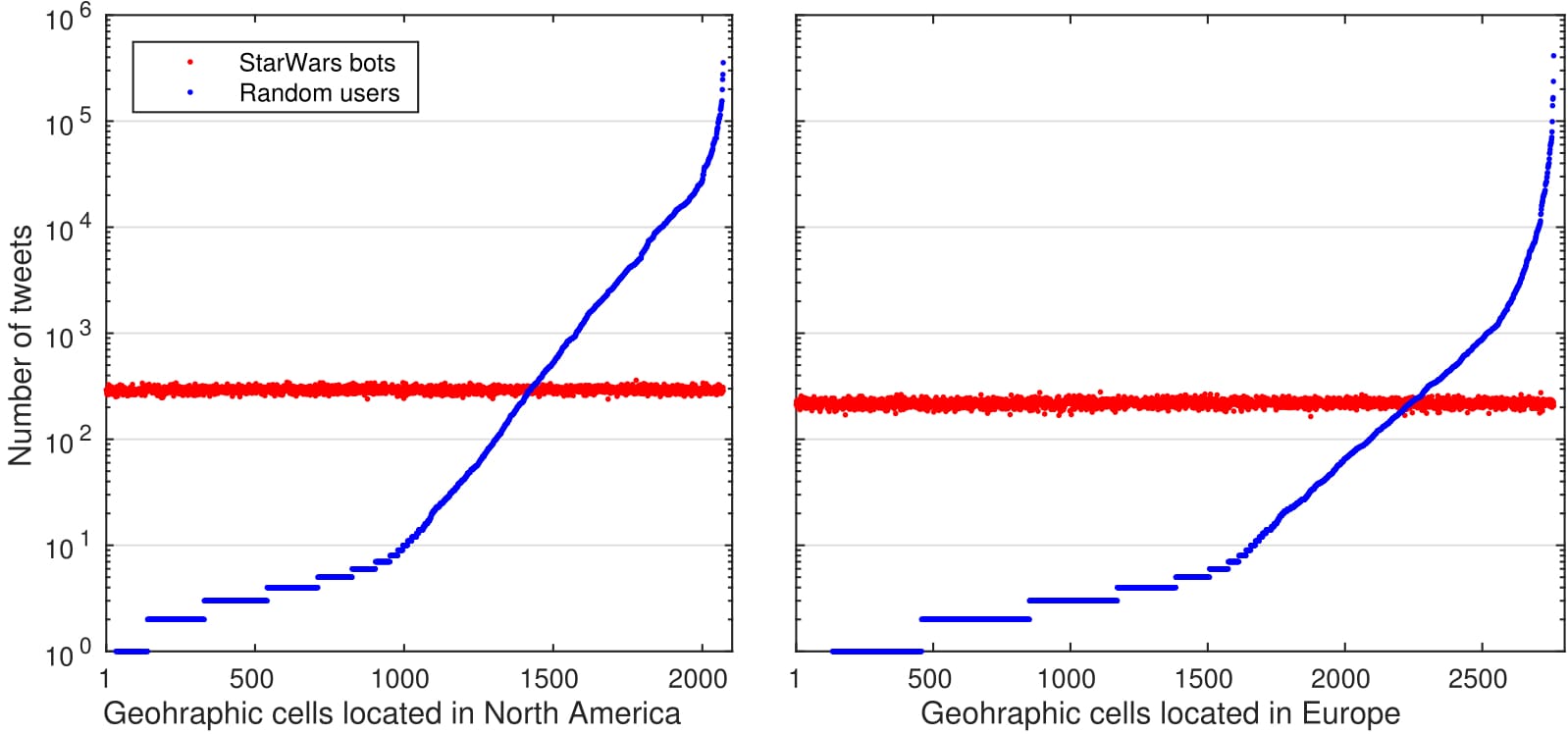}
	\caption{Number of tweets created by the Star Wars bots and the random users in  geographic cells inside the two rectangles. The cells in each rectangle are ordered with increasing number of tweets created by the random users.
	}
	\label{fig: CellDistribution}
\end{figure*}

As expected, Figure \ref{fig: Map bots} shows that the 1.2 million  tweets  are {\em uniformly} distributed in the two rectangles. The uniform distribution is also confirmed by examining the number of tweets in each cell as shown in Figure \ref{fig: CellDistribution}.
These unusual properties strongly indicate an artificial control by computer programmes.

\begin{figure}

 	\includegraphics[width=\columnwidth]{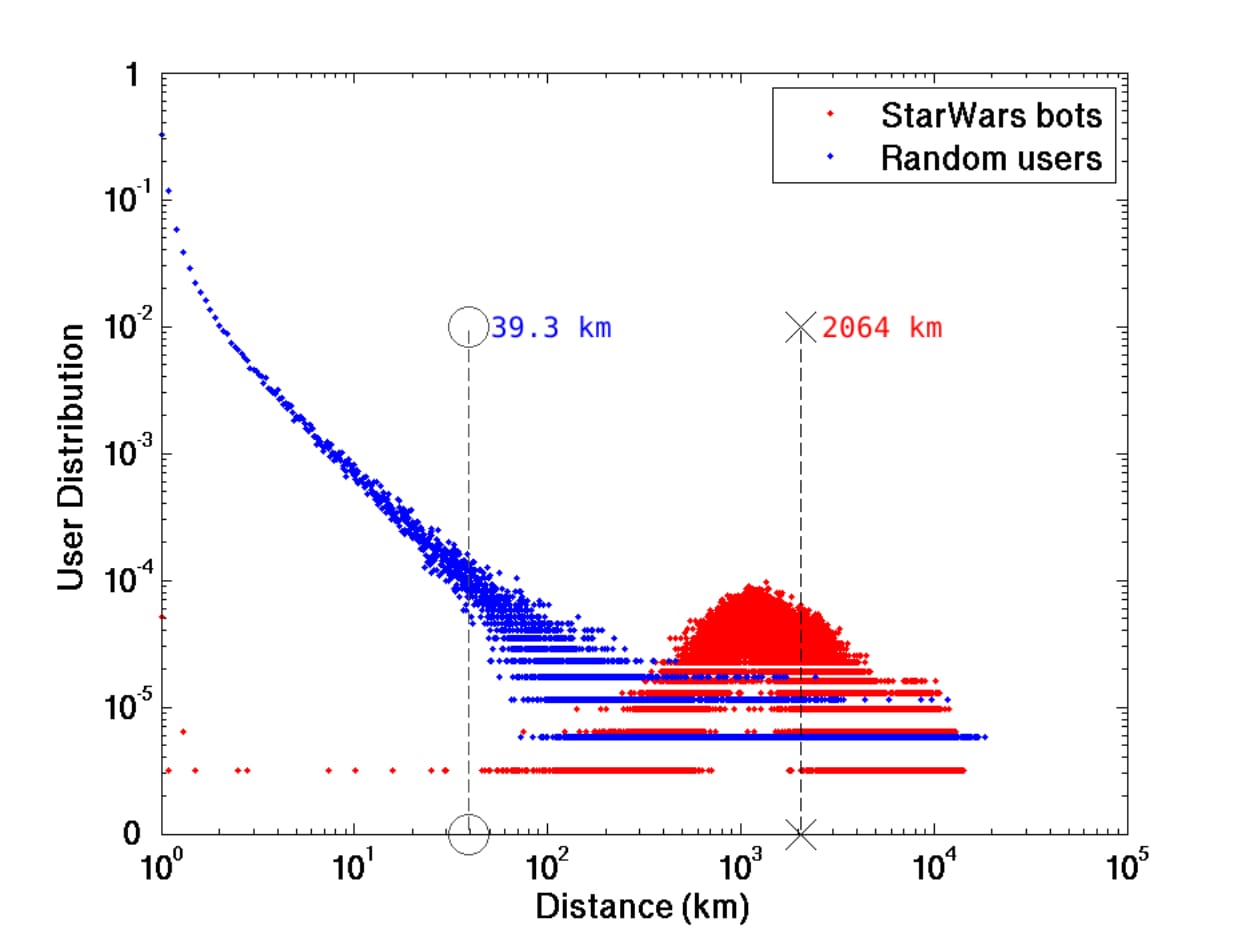}
 \caption{Distribution of the Star Wars bots and the random users as a function of the average Haversine  distance between consecutive location-tagged tweets. The two  vertical dash lines show the   average Haversine distance for all the bots and all the random users, respectively.  
 	}
 	\label{fig: DistanceDistribution}
 \end{figure}

We also calculated the average Haversine distance\footnote{Defined as the  surface distance between two locations taking into account of the Earth's near-sphere shape \cite{robusto_cosine-haversine_1957}.}   between two consecutive location-tagged tweets of a user.  
Figure \ref{fig: DistanceDistribution}  shows that the distribution of this quantity for the random users follows a power-law, with an average distance of less than 40 km. In sharp contrast, the distribution for the Star Wars bots resembles a bell curve, with an  average distance of 2,064km, which is about the distance between  centres of the two rectangles. 

This suggests that when a Star Wars bot  fakes its tweet locations, it firstly randomly chooses a continent, either Europe or North America, with equal probability; and  then randomly generates a pair of  latitude and longitude values within the ranges of that region.  As such, many consecutive tweets are located in different continents.  

\subsection{Connectivity properties}

After collecting all friends and followers of the Star Wars bots, we noticed the tell-tale signs of bot activity, i.e. the bots disproportionately follow other bots in the same botnet. When a user A follows a user B, we define A as a {\em follower} of B; and B as a {\em friend} of A.  Such following relationship can be represented as a directed link A-->B, which is an incoming link of B  or an outgoing link of A.  The Star Wars bots have in total 733,000 followers and 1,817,000 friends. 

\begin{figure}
	\centering
		\includegraphics{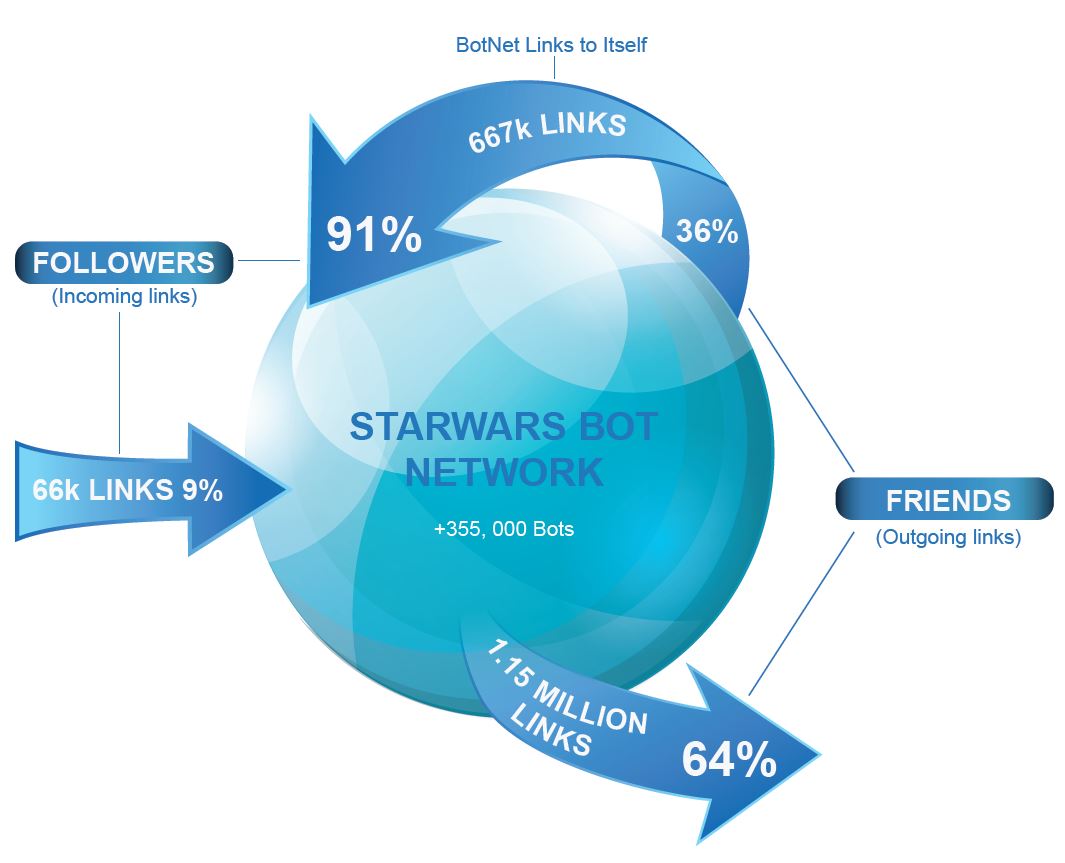}
		
		\caption{ Star Wars botnet's incoming links come mostly from the botnet itself }	\label{fig : SW_Follows}
		\end{figure}

As illustrated in Figure \ref{fig : SW_Follows}, 91\% of the bots' incoming links come from the botnet itself; in other words, most of the bots' followers are other Star Wars bots.  

Of the 1.8 million  outgoing links,  1/3 follow  other Star Wars bots and   2/3 follow users outside of the botnet. Additionally, Star Wars bots follow a few users disproportionately, this means many bots follow the same users outside of the botnet. Table \ref{tab: userslinkedtobots} shows some examples of users heavily followed by the botnet, but which are not part of it.

\section{Discussion}

\subsection{Creation of the Star Wars  botnet}

The above analysis reveals many details on the design and creation of the Star Wars bots.

Firstly, the user IDs of the  Star Wars bots are  between $ 1.5\times10^9$ and $1.6\times10^9$. From this we can infer that the Star Wars botnet was created between 20 June - 14 July 2013, because that was the time period Twitter assigned user IDs from that range to new users, we naturally double checked against the actual date from the API and found this to be true.   

Secondly, each Star Wars bot only created a few tweets, all of which were random quotations from one of a series of Star Wars novels. Some tweets have random hashtags inserted. 

Thirdly, half of the tweets were tagged with a faked location, which was a random location within one of the two rectangles.  

Fourthly, the bots only claimed Windows phone as tweet source. And they did not retweet or mention other users. 

\subsection{Reflection on the discovery   }

We discovered the Star Wars bots by accident.  
The bots tagged their tweets with random locations in North America and Europe   as  a deliberated effort to make their tweets look more real. But this  trick  backfired -- the fake locations  when plotted on a map were completely abnormal. {It's important to note that this anomaly  could only be noticed  by a human looking at the map. A computer algorithm would have  a hard time to realise the anomaly, especially when the anomaly only associates with areas with low density distribution of tweets.}

It was also a  `mistake' for the bots to use the Star Wars novels as the sole source of their tweets.  So when we studied the abnormal tweets, the Star Wars theme was easy to see. It was also this  feature  that allowed us to use a simple classifier to accurately identify all Star War bots.  
 
In addition, the fact that all of the bots were created during a relatively small period of time (and thus were given user IDs within a narrow range), gave us the additional convenience to test only users registered in that period. 

\subsection{`Success' of the Star Wars botnet   }

There have been a lot of efforts to detect and remove bots from Twitter. How can such a huge botnet,  created in 2013,  remain  hidden until now?  
 
Firstly it seems the Star Wars bots  were deliberately designed  to keep a low profile. They tweeted a few times, not too many, not too few. They avoided doing anything special. 

Secondly they only tweeted random quotations from novels. This helped the tweets appear like using real human's language. This invalidates many bot detection methods  based on detecting language created by machines. 

Thirdly the bots were carefully designed to have `normal' user profiles. We observed that some of the bots even have profile pictures. This invalidates the detection methods based on profile analysis. 

\begin{figure} 
	
	\includegraphics[width=\columnwidth]{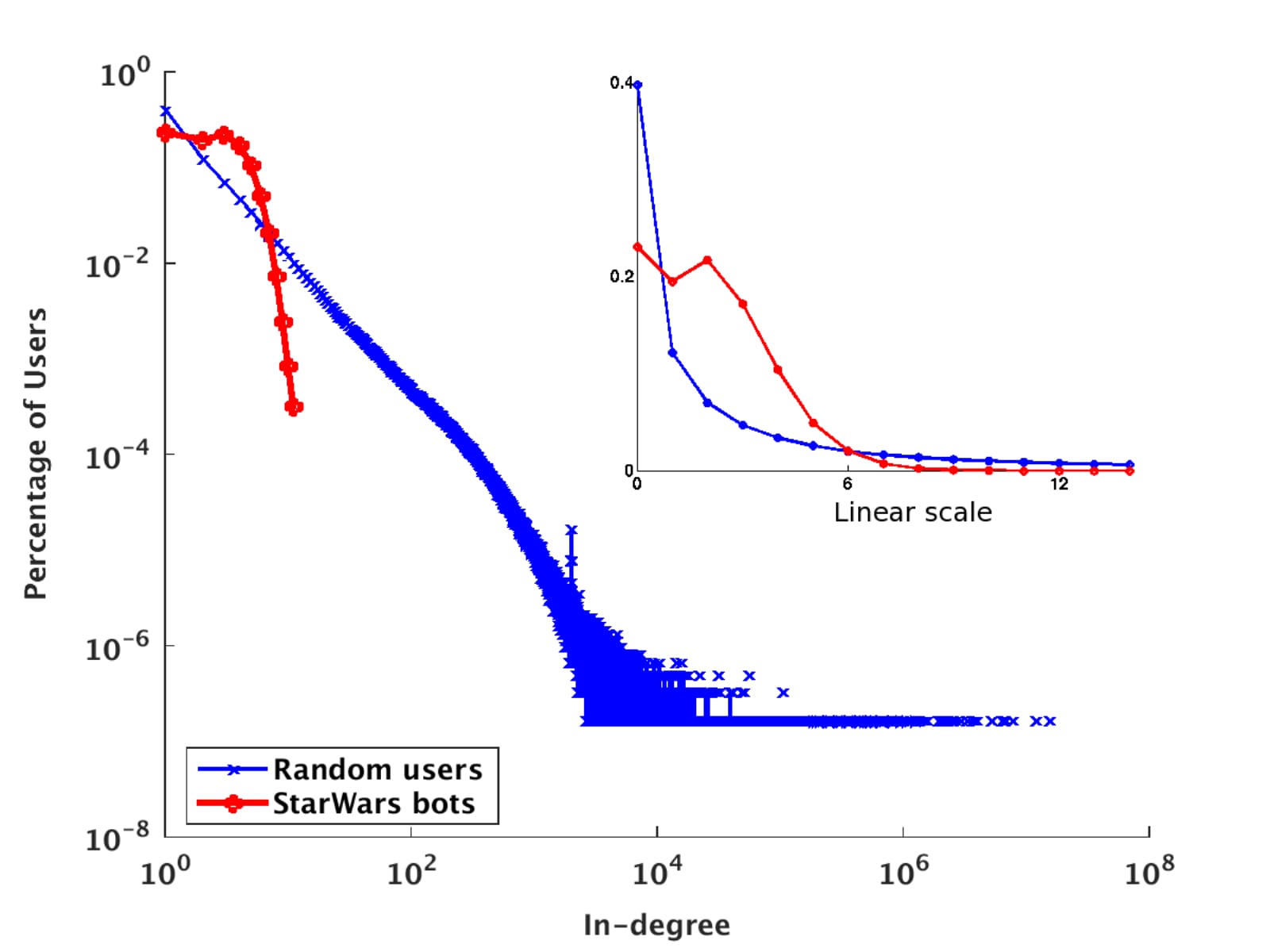}
	
	\includegraphics[width=\columnwidth]{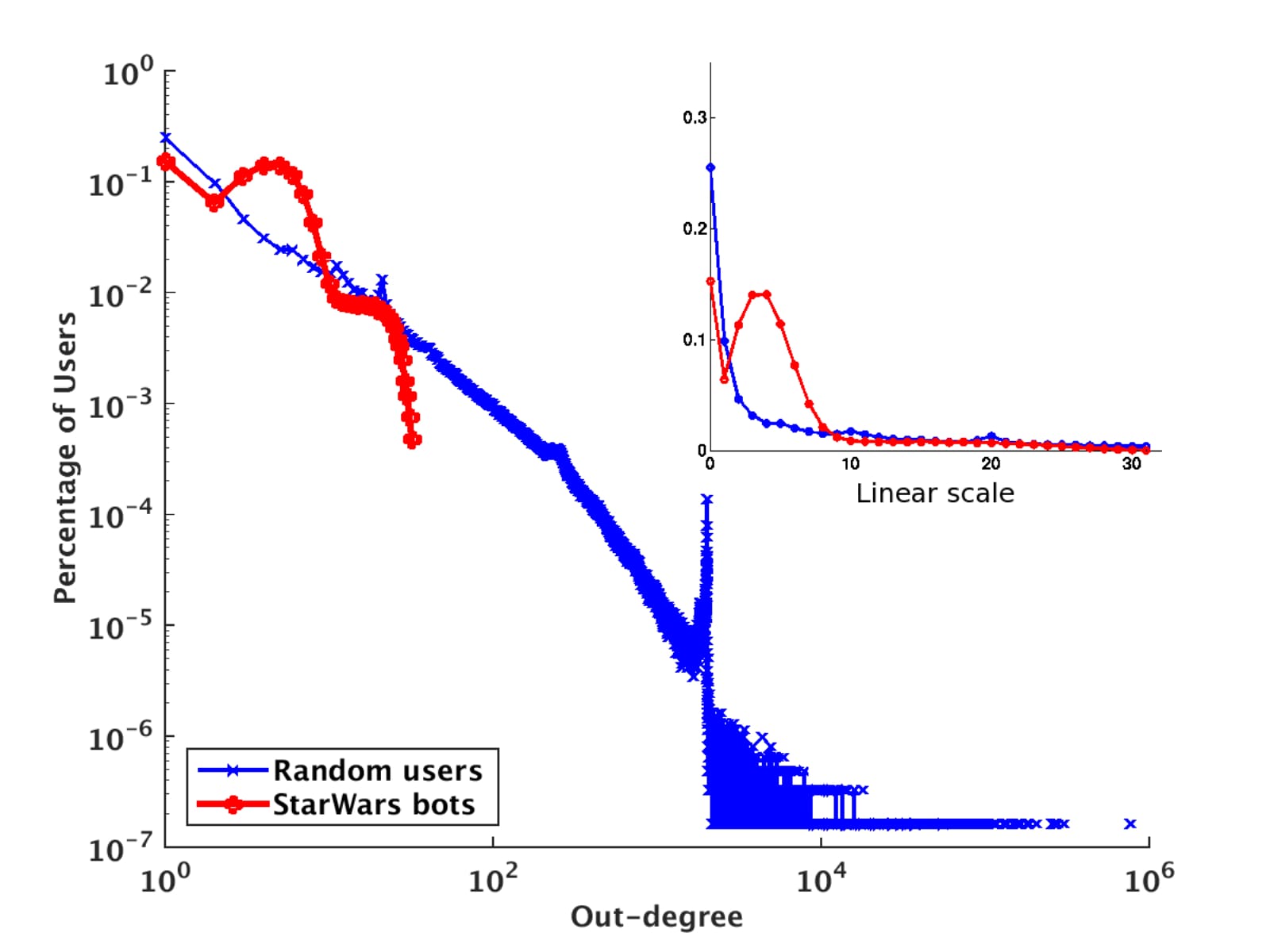}
	
	\caption{Distributions of in-degree (followers) and out-degree (friends or followees) for the Star Wars bots and the random users. The insets show the same distributions in  linear scale. }	\label{fig : In_degree}
\end{figure}

Finally and more importantly, it seems the   Star Wars bots were deliberately designed to circumvent many of the heuristics  underlying    bot detection methods.  For example,  contrary to popular assumptions on bots, the Star Wars bots do not have any URLs in their tweets, they never mention or reply to other users, and they only follow a small number of friends in comparison to random users (see Figure \ref{fig : In_degree}).

\subsection{Critical reflection on Twitter bot detection}

It is likely that botmasters are keeping their eyes on the latest developments on bot detection techniques.  
For example, after reading this paper, bot designers may learn a lesson that there is no need to tag tweets with locations at all, because most users do not. Also, quotations should use diverse  sources, such as a random collection of novels, magazines or social media postings. 

Future botnets could easily be  programmed to avoid  such design  `mistakes'. 
Indeed, future bots can learn lessons from all previous detection methods, and it may    become more and more difficult to detect future botnets. 
%
So far there is no known feature that   can be used to distinguish bots from real users that cannot be deliberately modified by bots. 

Given the challenge ahead, it is time to reflect on  some of the problems in the bot detection research area. For example, many detection methods are based on assumptions about bots, but many of such assumptions are not sufficiently supported by data evidence. 

Secondly, recent efforts on Twitter bot detections are increasingly focused on exploring  machine learning techniques, where bot detection becomes merely an `application' of the  techniques. There are much fewer efforts in collecting   ground truth data and  analysing real bot behaviours.  
In this work we  used   one of the most basic machine learning classifiers, however we uncovered a botnet far larger than most of  previous research efforts.

Finally, some researchers  are keen to seek for a `general' method that is able to detect all bots, but they fail to explain why vastly different types of bots or botnets should share any common features at all. Indeed, existing works aiming for a general detection method have only detected a  small number of bots. Such a `general' approach will become more problematic as future new bots will be designed to be different from any bots that have  been detected and reported.

\subsection{Potential threats of dormant Twitter bots }

The low tweet count and   the long period of inactivity of the Star Wars bots might look like a reason to think they are harmless. 
The years-long inactivity may make one wonder whether the bots have been forgotten by their master, or the access credentials were lost,  or the bots were created only for fun?
However, it takes effort and resources to create such a large number of bots and it is unlikely they have been forgotten, especially when they are actually very useful and valuable. 

What if the master of the botnet deliberately make them dormant with a low profile? If so, not only the bots have successfully defeated  previous detection efforts, but also they can in theory  
 pose   all of the threats discussed in Section \ref{section:threats}, including spam, fake trending topics, opinion manipulation, and astroturfing. Threats could be enacted whenever the botmaster chooses to re-activate the botnet.


It is known \cite{thomas_trafficking_2013} that pre-aged bots are sold at a premium on the black market.  
This means the Star Wars bots are perfectly  suited to be sold as fake followers because they are already four years old and therefore  more 'valuable'.  Furthermore, they have a few tweets and a few connections each, making them seem more human than an "egg" account created the day before it is sold.

For example, as shown in Table \ref{tab: userslinkedtobots}, we found that there are a few   users (who are not members of the Star Wars botnet) are followed by  large numbers of the Star Wars bots.   It is likely these bots have  already been sold as fake followers to those users.   

\begin{table}
	\centering
	\begin{tabular}{|c|c|c|c|}
		\hline 
		\textbf{Twitter user ID} &\textbf{Number of Star Wars bot followers}\tabularnewline
		\hline 
		\hline 
		 19196415 & 15,955\tabularnewline
		\hline 
	  14230524 & 14,315\tabularnewline
		\hline 
		 39922782 & 13,018\tabularnewline
		\hline 
		  29547045 & 11,946\tabularnewline
		\hline 
	\end{tabular}
	\caption{Examples of  large numbers of Star Wars bots following the same Twitter users outside of the Star Wars botnet}
	\label{tab: userslinkedtobots}
\end{table}
 
 In recent years, fake online user accounts, including Twitter bots, have been increasingly used for online advertisement fraud with billions of dollars on stake \cite{_fake_traffic}, where computers log into fake accounts and then browser webpages   to generate fake impressions and clicks. In this case, even if the bots are being used daily, we will not see any tweeting activities.   
 
\section{Conclusion}

The discovery of the Star Wars botnet provides a valuable source of ground truth data for  research on Twitter bots.  
It is very large,  many times larger than publicly available datasets. More importantly, it contains only a single botnet -- and the whole of the botnet, which allows us to reveal the details on how the botnet was designed and created. 
As of this writing, the Star Wars bots are still alive on Twitter, which provides researchers a rare opportunity to monitor and study the bots. 
The more we know and understand the design and behaviour of bots, the more likely we will be able to propose effective methods to uncover and fight against them.  

The data of the Star Wars bots are available for researchers on request. Researchers can also collected the bots by following  details in this paper. 
Recently we discovered a special property of the Star Wars bots that has never been reported before. Based on this knowledge, we discovered another every larger Twitter botnet with more than 500k bots. This will be reported in a separate paper.





\bibliographystyle{IEEEtran}


\end{document}